\def\rfr#1{eq. (\ref{#1})}

\def\Rfr#1{Eq. (\ref{#1})}

\def\dert#1#2{\frac{{{d}}{#1}}{{{d}}{#2}}}              

\def\dert#1#2{\frac{{{d}}{#1}}{{{d}}{#2}}}              

\def\bar{\begin{eqnarray}}
\def\ear{\end{eqnarray}}

\def\eqi{\begin{equation}}
\def\eqf{\end{equation}}
\def\eqia{\begin{eqnarray}}
\def\eqfa{\end{eqnarray}}
\def\rp#1#2{\frac{#1}{#2}}
\def\ct#1{\cite{#1}}
\def\lb#1{\label{#1}}

\def\bds#1{\boldsymbol{#1}}

\documentclass[11pt]{article}
\usepackage{amsmath,amsthm,amscd,amssymb}
\usepackage{latexsym}
\usepackage{graphicx,epsfig}

\begin{document}

\noindent{\bf \LARGE{MOND orbits in the  Oort cloud}}
\\
\\
\\
{L. Iorio$^{\ast}$}\\
{\it $^{\ast}$Ministero dell'Istruzione, dell'Universit\`{a} e della Ricerca (M.I.U.R.),\\
Fellow of the Royal Astronomical Society (F.R.A.S.).\\
 Viale Unit\`{a} di Italia 68, 70125, Bari (BA), Italy. Tel. 0039 328 6128815. email: lorenzo.iorio$@$libero.it }

\vspace{4mm}

\begin{abstract}
We numerically investigate the features of typical orbits occurring in the Oort cloud ($r\approx 50-150$ kAU) in the low-acceleration regime of the MOdified Newtonian Dynamics (MOND). We fully take into account the so-called External Field Effect (EFE) because the solar system is embedded in the Milky Way. In the framework of MOND this does matter since the gravitational acceleration of Galactic origin felt by the solar system is of the same order of magnitude of the characteristic MOND acceleration scale $A_0\approx 10^{-10}$ m s$^{-2}$. We use three different forms of the MOND interpolating function $\mu(x)$, two different values for the Galactic field at the Sun's location and different initial conditions corresponding to plausible Keplerian ellipses in the Oort cloud. We find that MOND produces highly distorted trajectories with respect to the Newtonian case, especially for very eccentric orbits.  It turns out that  the shape of the MOND orbits strongly depend on the initial conditions. For  particular initial state vectors, the MOND paths in the ecliptic plane get shrunk extending over much smaller spatial regions than in the Newtonian case, and experience high frequency variations over one Keplerian orbital period.  Ecliptic orbits with different initial conditions  and  nearly polar orbits are quite different getting distorted as well, but
they occupy more extended spatial regions. These facts may have consequences on the composition and the dynamical history of the Oort cloud which are difficult to predict in detail; certainly, the MOND picture of the Oort region is quite different from the Newtonian one exhibiting no regularities.
\end{abstract}



\section{Introduction}
In many astrophysical systems like, e.g., spiral galaxies and clusters of galaxies a discrepancy between the observed kinematics of their exterior parts and the predicted one on the basis of the Newtonian dynamics and the matter detected from the emitted electromagnetic radiation (visible stars and gas clouds) was present  since the pioneering studies, in recent times,  by Bosma \cite{Bos} and Rubin and coworkers \cite{Rub} on spiral galaxies. More precisely, such an effect shows up in the galactic  velocity rotation curves  whose typical pattern after a few kpc from the center differs from the Keplerian $1/\sqrt{r}$ fall-off expected from the usual dynamics applied to the electromagnetically-observed matter.

As a possible solution of this puzzle, the existence of non-baryonic, weakly-interacting Cold Dark (in the sense that its existence is indirectly inferred only from its gravitational action, not from emitted electromagnetic radiation) Matter (CDM) was proposed to reconcile the predictions with the observations \ct{Rub83} in the framework of the standard gravitational physics.

Oppositely, it was postulated that the Newtonian laws of gravitation may have to be modified on certain acceleration scales to correctly account for the observed anomalous kinematics of such astrophysical systems without resorting to still undetected exotic forms of matter.
One of the most phenomenologically successful modification of the inverse-square Newtonian law, mainly with respect to spiral galaxies, is the MOdified Newtonian Dynamics (MOND) \ct{Mil83a,Mil83b,Mil83c} which  postulates that
for systems  experiencing total gravitational accelerations  $A < A_0$, with \ct{Bege} \eqi A_0= (1.2\pm 0.27)\times 10^{-10} \ {\rm m\ s}^{-2},\eqf
the overall gravitational acceleration felt gets modified according to \eqi \bds A\rightarrow \bds A_{\rm MOND}=-\rp{\sqrt{A_0GM}}{r}\bds{\hat{r}}.\lb{MOND}\eqf   More generally, it holds\footnote{Actually, \rfr{appromond1} is exactly valid for isolated mass distributions endowed with particular symmetries.} \eqi A = \rp{A_{\rm Newton}}{\mu(x)},\ x\equiv\rp{A}{A_0};\lb{appromond1}\eqf
$\mu(x)\rightarrow 1$ for $x\gg 1$, i.e. for large accelerations (with respect to $A_0$), while $\mu(x)\rightarrow x$ yielding \rfr{MOND} for $x\ll 1$, i.e. for small accelerations.
The most widely used forms for the interpolating function $\mu$ are \cite{Fam,joint}
\begin{eqnarray}
  \mu_1(x) &=& \rp{x}{1+x},\lb{mu1}\\
  \mu_2(x) &=& \rp{x}{(1+x^2)^{1/2}}.\lb{mu2}
\end{eqnarray}
Such forms, and also others, as we will see later, can be reduced to the following high-acceleration limit ($x\gg 1$)
\eqi \mu\approx 1-k_0 x^{-\alpha}.\eqf Indeed, \rfr{mu1} corresponds to $k_0=1$, $\alpha=1$, while \rfr{mu2} corresponds to $k_0=1/2$ and $\alpha=2$.
It recently turned out that the simpler form of \rfr{mu1} yields  better results in fitting the terminal velocity curve of the Milky Way, the rotation curve of the standard external galaxy NGC 3198 \cite{Fam,kazzo,scassa} and of a sample of 17 high
surface brightness, early-type disc galaxies \cite{Noor}.
\Rfr{appromond1} strictly holds for co-planar, spherically and axially symmetric isolated mass distributions \cite{BradaMil95}; otherwise, for generic mass densities the full  modified (non-relativistic) Poisson equation\footnote{By posing $\bds A\doteq -\bds\nabla U$, \rfr{poiss} implies that, in general, $\mu(A/A_0)\bds A=\bds A_{\rm Newton}+\bds\nabla\bds\times\bds h$; it can be shown \cite{joint} that at great distances from an isolated matter distribution with mass $M$ it holds $\mu(A/A_0)\bds A=-(GM/r^2)\bds{\hat r}+\mathcal{O}(r^{-3})$.}   \cite{joint}
\eqi \bds\nabla\bds\cdot\left[\mu\left(\rp{|\bds\nabla U|}{A_0}\right)\bds\nabla U\right]=4\pi G\rho,\lb{poiss}\eqf where $U$ is the gravitational potential, $G$ is the Newtonian constant of gravitation and $\rho$ is the matter density generating $U$,
must be used.

Attempts to yield a physical foundation to MOND, especially in terms of a relativistic covariant theory, can be be found in, e.g., \ct{Bek,Far,Zha07}; for recent reviews of various aspects of the MOND paradigm, see \ct{San02,Bek06,Mil08}.


 After setting the theoretical background which we will use in the rest of the paper,  we will explore the strong MONDian regime in the remote ($r=50-150$ kAU) periphery of the solar system, where the Oort cloud \cite{Oo}, populated by a huge number of small bodies moving along very eccentric and inclined to the ecliptic orbits, is supposed to exist; for preliminary investigations on such a topic, see \cite{Mil83a,Milpazzo}. In particular, we will numerically investigate the  modifications that MOND would induce on the Newtonian orbits of a test particle moving in such a region.
 We will also have to consider the subtle External Field Effect (EFE) which, in MOND, takes into account the influence of the Galactic field in the dynamics of the bodies of the solar system giving rise, in the regions in which $x\ll 1$, to a total gravitational acceleration quite different from the scheme consisting of the Newtonian one plus some perturbative correction(s) to it.

For other works  on MOND in the inner regions of the solar system, see Ref. \ct{Mil83a,Tal,Ser,Magu,San06,Yu1,Ior,Yu2,Mil09,Ior09}.
\section{The External Field Effect in MOND}
In the framework of MOND, the internal dynamics of a
gravitating system s embedded in a larger one S is affected
by the external background field $\bds E$ of S even if
it is constant and uniform, thus implying a violation
of the Strong Equivalence Principle: it is the so-called
External Field Effect (EFE). In the case of the solar
system, $E$ would be $A_{\rm cen}\approx 10^{-10}$ m s$^{-2}$ because of its
motion through the Milky Way \cite{Mil83a,joint,Milpazzo}.

Clarifying EFE's concrete effects on the orbital dynamics of s is
not an easy task also because of some possible misunderstandings due to the intrinsic difficulty of the subject. The existence in the inner regions of the solar system of its simplest form\footnote{This was, at least, the view about EFE in the solar system of some researchers active in the field.}, i.e. a constant and uniform vector field $\bds E$ added to the standard Newtonian monopole with $E\approx 10^{-10}$ m s$^{-2}$, has been ruled out in Ref.~\cite{IorApSS} by comparing analytical calculation with the estimated corrections $\Delta\dot\varpi$ to the standard perihelion precessions of some planets determined with the  EPM ephemerides by E.V. Pitjeva (RAS, IAA) \cite{Pit05}; it turned out that the upper bound on an anomalous acceleration with the characteristics of $\bds E$ is of the order of $10^{-14}$ m s$^{-2}$. Such a result has been recently confirmed by W.M. Folkner \cite{Folk}, although in a different context\footnote{Actually, he considered an extra-acceleration directed radially towards the Sun.}, with a modified version of the latest  DE ephemerides by NASA JPL including a constant and uniform extra-acceleration with which he fitted long observational data records. Milgrom in Ref.~\cite{Mil09} has put forth a different form for the planetary EFE yielding an additional quadrupolar extra-acceleration which mimics the action of distant mass; some consequences have been investigated in Ref.~\cite{Ior09}.

Moving to the remote periphery of the solar system, which is our present target, let us define the following quantities
\begin{eqnarray}
  \eta &\doteq& \rp{A_{\rm cen}}{A_0}\gtrsim 1, \\
  r_t &\doteq& \sqrt{\rp{GM}{A_0}} = 6.833\ {\rm kAU},\\
  L &\doteq& \rp{x}{\mu}\left(\dert \mu x\right), \\
  \mu_g &\doteq& \mu(\eta), \\
  L_g &\doteq& L(\eta).
\end{eqnarray}
In the weak acceleration regime, for
\eqi r\gg {r_t}\eta^{-1/2},\lb{condicio}\eqf which is fully satisfied in the Oort cloud, the  action of EFE is different, so that the total barycentric acceleration felt by an Oort comet is\footnote{It comes from \rfr{poiss} by imposing a particular boundary condition, i.e. that for $r\rightarrow\infty$, $\bds\nabla U\rightarrow-\bds E$, and by assuming $A/E\ll 1$ \cite{joint}.} \cite{joint,Mil09}
\eqi \bds A = -\rp{GM}{\mu_g(1+L_g)^{1/2}}\left(\rp{x^2}{1+L_g} + y^2 + z^2\right)^{-\rp{3}{2}}\left(
                                                                                           \begin{array}{c}
                                                                                             \rp{x}{1+L_g} \\
                                                                                             y \\
                                                                                             z \\
                                                                                           \end{array}
                                                                                         \right).\lb{efaccel}
\eqf
It must be stressed that, contrary to what it could be thought at first sight, \rfr{efaccel} is not a small correction to the Newtonian monopole $\bds A_{\rm Newton}=-(GM/r^2)\bds{\hat{r}}$ to be tackled with the usual perturbative techniques like, e.g., the Gauss equations for the variation of the elements. Indeed, \rfr{efaccel} must be treated as a whole because it is the total gravitational pull felt by a body in the weak acceleration regime according to MOND; it fully embeds EFE which cannot be disentangled. The only consistent alternative to the MONDian acceleration of\footnote{Given the scenario considered here, i.e. the solar system embedded in the Milky Way, it would be inconsistent to consider the simple form of the MOND acceleration of \rfr{MOND}. Indeed, it does not encompass EFE, holding only for isolated systems or when $E/A_O\ll 1$ like, e.g., in the case of some satellites of the Milky Way embedded in the external field of M31 Andromeda.} \rfr{efaccel} is the standard Newtonian term.
It may be interesting to show, from a purely phenomenological point of view, that applying \rfr{efaccel} to the major bodies of the solar system in its planetary regions yields absurd results. Indeed, by considering, e.g., the Earth, the discrepancy between \rfr{efaccel} and the standard Newtonian monopole turns out to be of the order of $2-8\times 10^{-4}$ m s$^{-2}$ which corresponds to an enormous shift of up to $1\times 10^{10}$ m in the heliocentric radial distance of our planet.

Note that, since the ecliptic longitude and latitude of the Galactic Center are about $\lambda_{\rm GC}\approx 180$ deg, $\beta_{\rm GC}\approx -6$ deg \cite{Ior09}, EFE is directed along the $X$ axis of the ICRF, i.e. the barycentric frame in which the motion of solar system's objects are usually studied.
Concerning $L$, we have\footnote{The form $\mu_{3/2}$ is in Ref.~\cite{Mil09}.}
\begin{eqnarray}
  \mu_1 = \rp{x}{1+x} &\rightarrow & L_1 =\rp{1}{1+x}, \\
  \mu_2 = \rp{x}{(1+x^2)^{1/2}} &\rightarrow & L_2 = \rp{1}{1+x^2},\\
  \mu_{3/2} = \rp{x}{(1+x^{3/2})^{2/3}}&\rightarrow & L_{3/2} = \rp{1}{1+x^{3/2}}.
\end{eqnarray}
The value of $\eta$ depends on the Galactic field at the solar system's location which can be obtained from its centrifugal acceleration
\eqi A_{\rm cen}=\rp{V^2}{R},\eqf where $V$ is the speed of the Local Standard of Rest (LSR) and $R=8.5$ kpc is the Galactocentric distance. The standard IAU value for  the LSR speed is $V=220$ km s$^{-1}$ \cite{iau}, but recent determinations  obtained with the Very Long Baseline Array and the Japanese VLBI Exploration of Radio Astronomy project yield a higher value: $V=254\pm 16$ km s$^{-1}$ \cite{Reid}.
Thus, $\eta$ ranges from 1.5 to 2.3.
\section{Orbits of Oort comets in MOND}
\subsection{Ecliptic orbits}
\begin{figure}[ht!]
 \centerline{\psfig{file=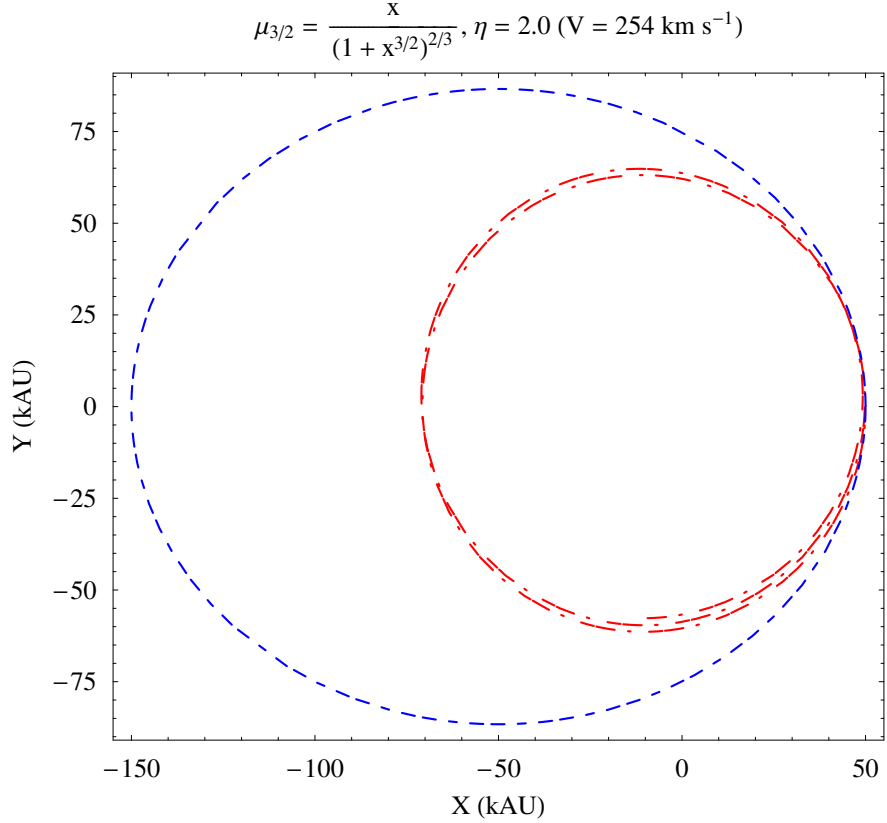,width=\columnwidth}}
\vspace*{8pt}
\caption{Numerically integrated orbits of an Oort comet with $a=100$ kAU, $e=0.5$, $P_{\rm b}=31.6$ Myr. Dashed blue line: Newton. Dash-dotted red line: MOND with $\mu_{3/2}$, $\eta=2.0$ ($V=254$ km s$^{-1}$). The initial conditions are $x_0=a(1-e), y_0=z_0=0,\dot x_0=0,\dot y_0=n a \sqrt{\rp{1+e}{1-e}},\dot z_0=0$. The time span of the integration is $- P_{\rm b}\leq t\leq 0$.}\label{uno}
\end{figure}

\begin{figure}[ht!]
 \centerline{\psfig{file=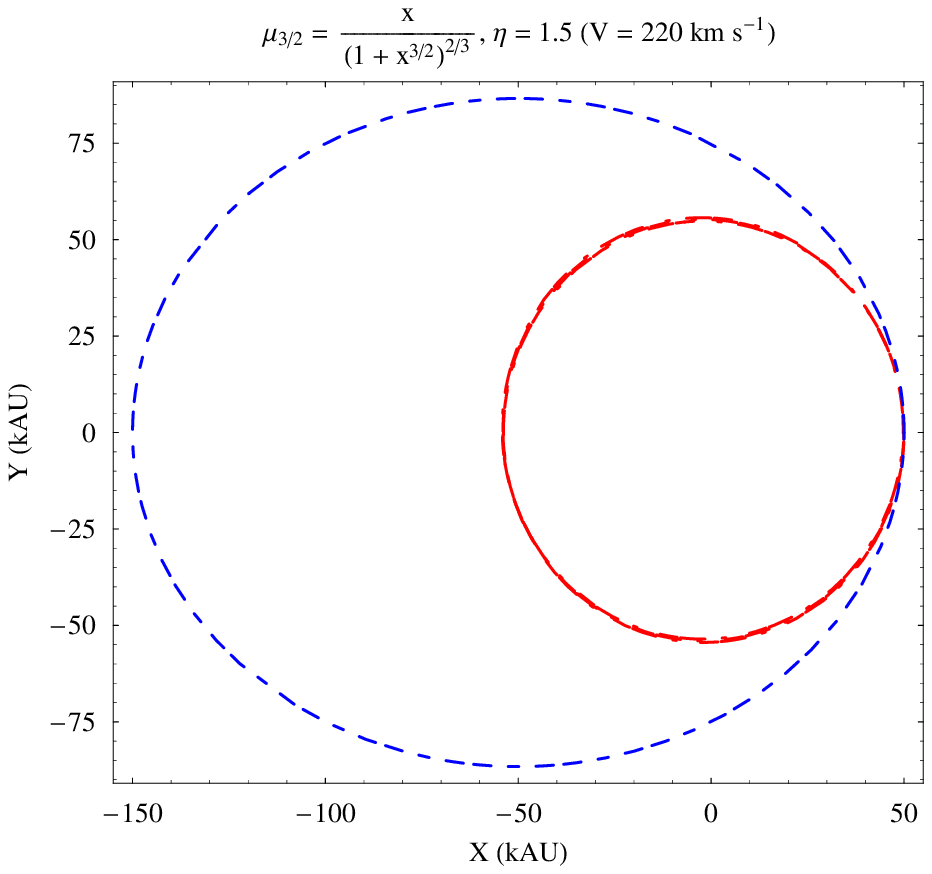,width=\columnwidth}}
\vspace*{8pt}
\caption{Numerically integrated orbits of an Oort comet with $a=100$ kAU, $e=0.5$, $P_{\rm b}=31.6$ Myr. Dashed blue line: Newton. Dash-dotted red line: MOND with $\mu_{3/2}$, $\eta=1.5$ ($V=220$ km s$^{-1}$). The initial conditions are $x_0=a(1-e), y_0=z_0=0,\dot x_0=0,\dot y_0=n a \sqrt{\rp{1+e}{1-e}},\dot z_0=0$. The time span of the integration is $- P_{\rm b}\leq t\leq 0$.}\label{due}
\end{figure}

\begin{figure}[ht!]
 \centerline{\psfig{file=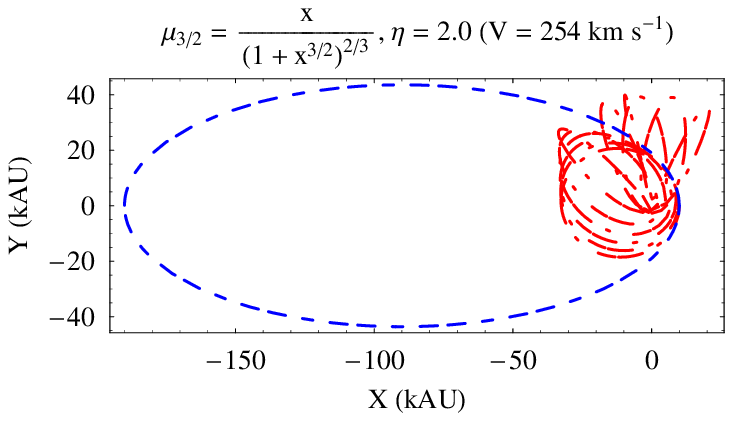,width=\columnwidth}}
\vspace*{8pt}
\caption{Numerically integrated orbits of an Oort comet with $a=100$ kAU, $e=0.9$, $P_{\rm b}=31.6$ Myr. Dashed blue line: Newton. Dash-dotted red line: MOND with $\mu_{3/2}$, $\eta=2.0$ ($V=254$ km s$^{-1}$). The initial conditions are $x_0=a(1-e), y_0=z_0=0,\dot x_0=0,\dot y_0=n a \sqrt{\rp{1+e}{1-e}},\dot z_0=0$. The time span of the integration is $- P_{\rm b}\leq t\leq 0$.}\label{tre}
\end{figure}

\begin{figure}[ht!]
 \centerline{\psfig{file=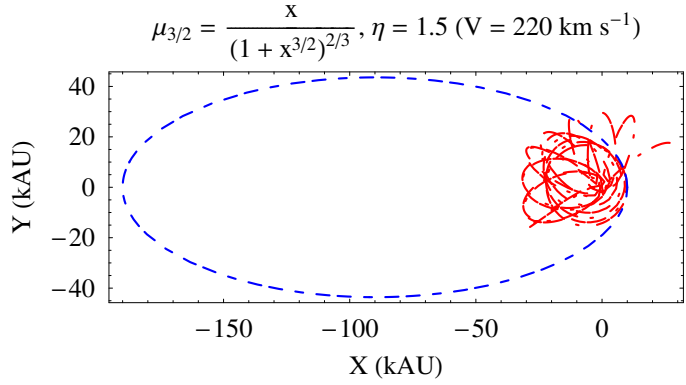,width=\columnwidth}}
\vspace*{8pt}
\caption{Numerically integrated orbits of an Oort comet with $a=100$ kAU, $e=0.9$, $P_{\rm b}=31.6$ Myr. Dashed blue line: Newton. Dash-dotted red line: MOND with $\mu_{3/2}$, $\eta=1.5$ ($V=220$ km s$^{-1}$). The initial conditions are $x_0=a(1-e), y_0=z_0=0,\dot x_0=0,\dot y_0=n a \sqrt{\rp{1+e}{1-e}},\dot z_0=0$. The time span of the integration is $- P_{\rm b}\leq t\leq 0$.}\label{quattro}
\end{figure}

\lb{eclitt}
We will, now, consider an Oort comet whose Newtonian orbit covers the entire extension of the Oort cloud. It has semimajor axis $a=100$ kAU and eccentricity $e=0.5$, so that its perihelion is 50 kAU and its aphelion is 150 kAU; for the sake of simplicity, we will assume it lies in the ecliptic plane. Its Newtonian orbital period is $P_{\rm b}=31.6$ Myr.
We will, first, use $\mu_{3/2}$. The value $\eta=2.0$, corresponding to $V=254\ {\rm km\ s^{-1}}$, yields
\begin{eqnarray}
  \mu_g &=& 0.82 \\
  L_g &=& 0.25
\end{eqnarray}
Figure \ref{uno} depicts the numerically integrated Newtonian (dashed blue line) and MONDian (dash-dotted red line) orbits for the same initial conditions. The integration has been performed backward in time over one (Keplerian) orbital period.
In Figure \ref{due} we show the case $\eta=1.5\ (V=220\ {\rm km\ s^{-1}})$ yielding
\begin{eqnarray}
  \mu_g &=& 0.75 \\
  L_g &=& 0.34.
\end{eqnarray}
The MOND trajectories are not closed
and the points of minimum and maximum distances from the Sun do not coincide with the Newtonian ones. In general, they do not show a regular pattern.
 Moreover, the MOND paths are much less spatially extended that the Newtonian ones; the overall shrinking of the orbit is more marked for the standard value of the LSR circular speed (Figure \ref{due}). Finally, the MOND trajectories experience high frequency variations during one (Keplerian) orbital period.
Such effects are particularly notable for highly elliptical Newtonian orbits, as shown by Figure \ref{tre}- Figure \ref{quattro} for $e=0.9$ and $-P_{\rm b}\leq t\leq 0$. In the Newtonian case, their perihelia are at 10 kAU, while the aphelia lie at 190 kAU, so that the condition of \rfr{condicio} is still well fulfilled.
The MOND orbits resemble involved clews comprised within 60 kAU $\times$ 60 kAU.

\begin{figure}[ht!]
 \centerline{\psfig{file=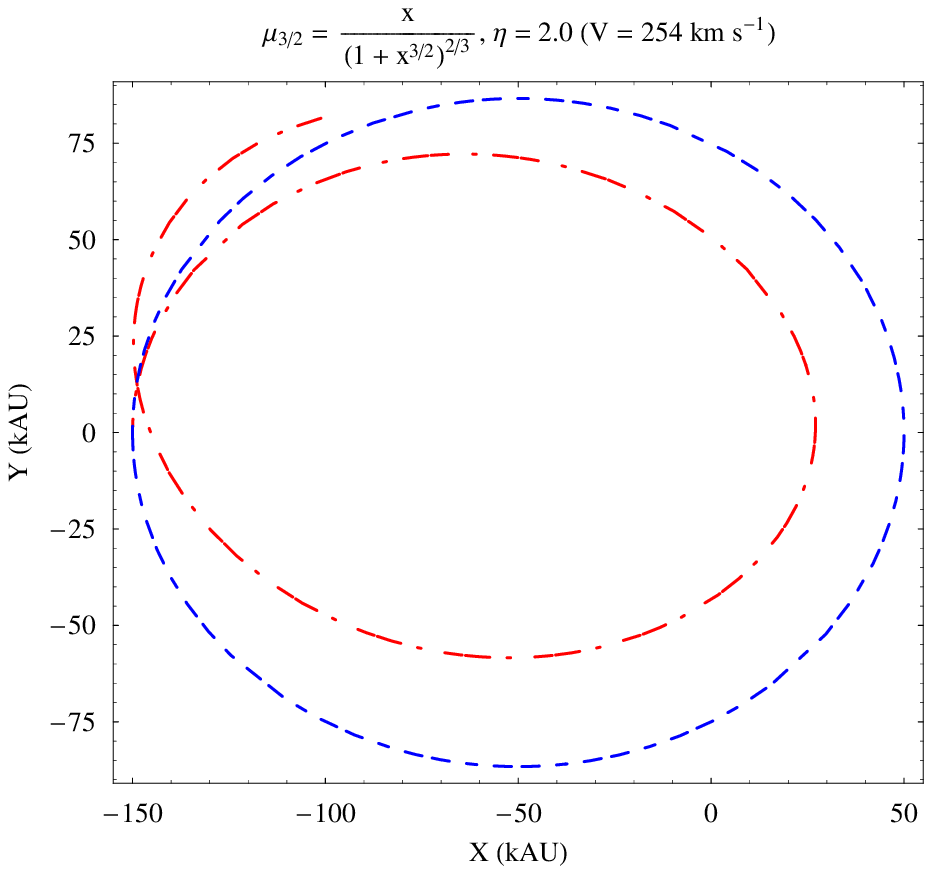,width=\columnwidth}}
\vspace*{8pt}
\caption{Numerically integrated orbits of an Oort comet with $a=100$ kAU, $e=0.5$, $P_{\rm b}=31.6$ Myr. Dashed blue line: Newton. Dash-dotted red line: MOND with $\mu_{3/2}$, $\eta=2.0$ ($V=254$ km s$^{-1}$). The initial conditions are $x_0=-a(1+e), y_0=z_0=0,\dot x_0=0,\dot y_0=-n a \sqrt{\rp{1-e}{1+e}},\dot z_0=0$. The time span of the integration is $-P_{\rm b}\leq t\leq 0$.}\label{afo1}
\end{figure}

\begin{figure}[ht!]
 \centerline{\psfig{file=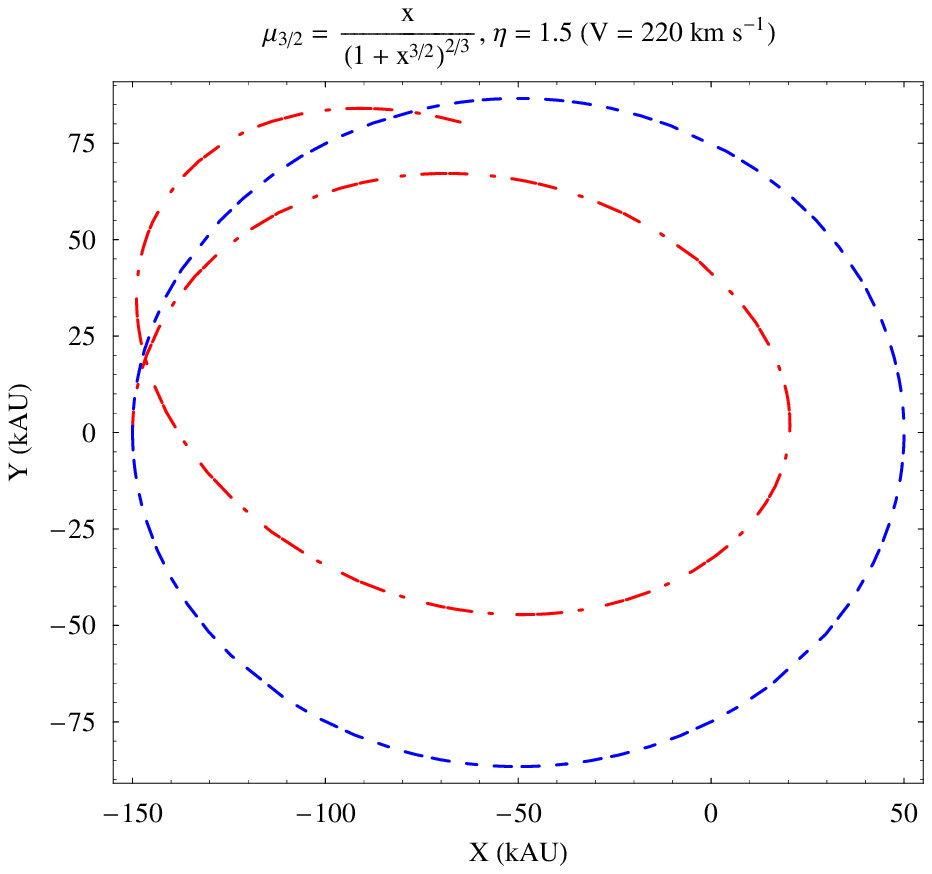,width=\columnwidth}}
\vspace*{8pt}
\caption{Numerically integrated orbits of an Oort comet with $a=100$ kAU, $e=0.5$, $P_{\rm b}=31.6$ Myr. Dashed blue line: Newton. Dash-dotted red line: MOND with $\mu_{3/2}$, $\eta=1.5$ ($V=220$ km s$^{-1}$). The initial conditions are $x_0=-a(1+e), y_0=z_0=0,\dot x_0=0,\dot y_0=-n a \sqrt{\rp{1-e}{1+e}},\dot z_0=0$. The time span of the integration is $- P_{\rm b}\leq t\leq 0$.}\label{afo2}
\end{figure}

\begin{figure}[ht!]
 \centerline{\psfig{file=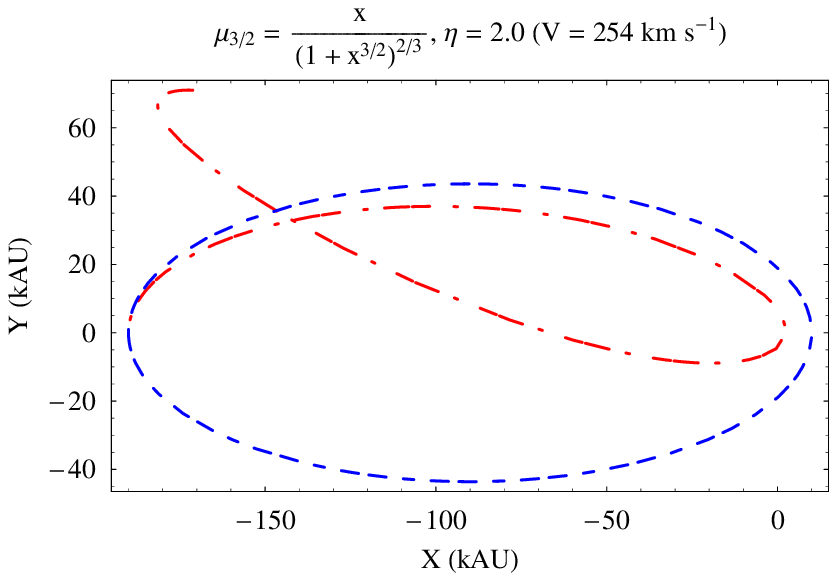,width=\columnwidth}}
\vspace*{8pt}
\caption{Numerically integrated orbits of an Oort comet with $a=100$ kAU, $e=0.9$, $P_{\rm b}=31.6$ Myr. Dashed blue line: Newton. Dash-dotted red line: MOND with $\mu_{3/2}$, $\eta=2.0$ ($V=254$ km s$^{-1}$). The initial conditions are $x_0=-a(1+e), y_0=z_0=0,\dot x_0=0,\dot y_0=-n a \sqrt{\rp{1-e}{1+e}},\dot z_0=0$. The time span of the integration is $- P_{\rm b}\leq t\leq 0$.}\label{afo3}
\end{figure}

\begin{figure}[ht!]
 \centerline{\psfig{file=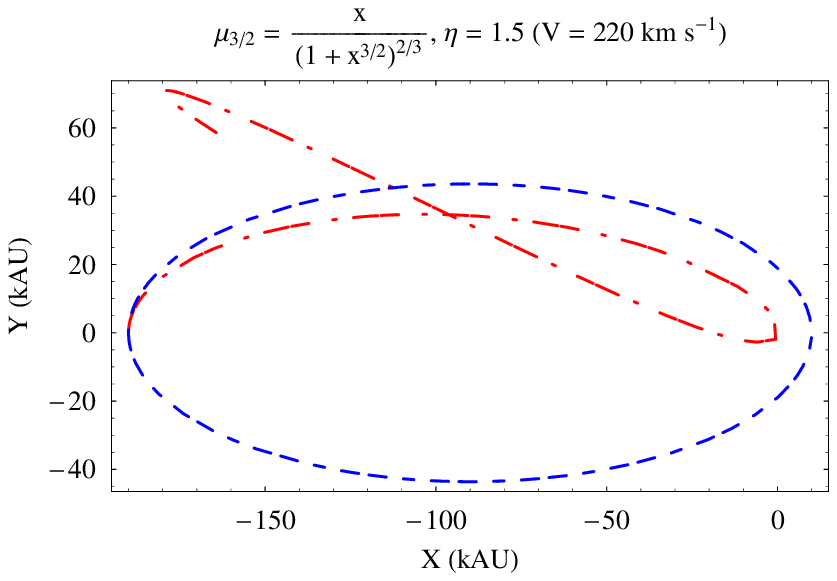,width=\columnwidth}}
\vspace*{8pt}
\caption{Numerically integrated orbits of an Oort comet with $a=100$ kAU, $e=0.9$, $P_{\rm b}=31.6$ Myr. Dashed blue line: Newton. Dash-dotted red line: MOND with $\mu_{3/2}$, $\eta=1.5$ ($V=220$ km s$^{-1}$). The initial conditions are $x_0=-a(1+e), y_0=z_0=0,\dot x_0=0,\dot y_0=-n a \sqrt{\rp{1-e}{1+e}},\dot z_0=0$. The time span of the integration is $- P_{\rm b}\leq t\leq 0$.}\label{afo4}
\end{figure}

At this point, it must be noted that in Figure \ref{uno}-Figure \ref{quattro} we used
the state vector at the perihelion as initial conditions for the numerical integrations.
We have to check if the behavior described is somewhat related to the peculiar initial conditions chosen. To this aim, in Figure \ref{afo1}-Figure \ref{afo2} we consider paths starting at the (Newtonian) aphelion for $e=0.5$,
while the case $e=0.9$ is depicted in Figure \ref{afo3}-Figure \ref{afo4}. Also in this cases we integrated the equations of motion backward in time over one (Keplerian) orbital period $P_{\rm b}$.
It can be noted that the MOND trajectories starting at the (Newtonian) aphelion are quite different with respect to those starting at the (Newtonian) perihelion depicted before. Again, they are not close and no regular patterns occur. In this case, the areas swept by the MOND paths are larger than in the previous case, and they occupy a large part of the Newtonian ones. For $e=0.9$ the MOND orbits approximately lie within 180 kAU $\times$ 60 kAU.

The case for $\mu_2$ is rather similar, so that we will not depict the related figures for saving space.

\begin{figure}[ht!]
 \centerline{\psfig{file=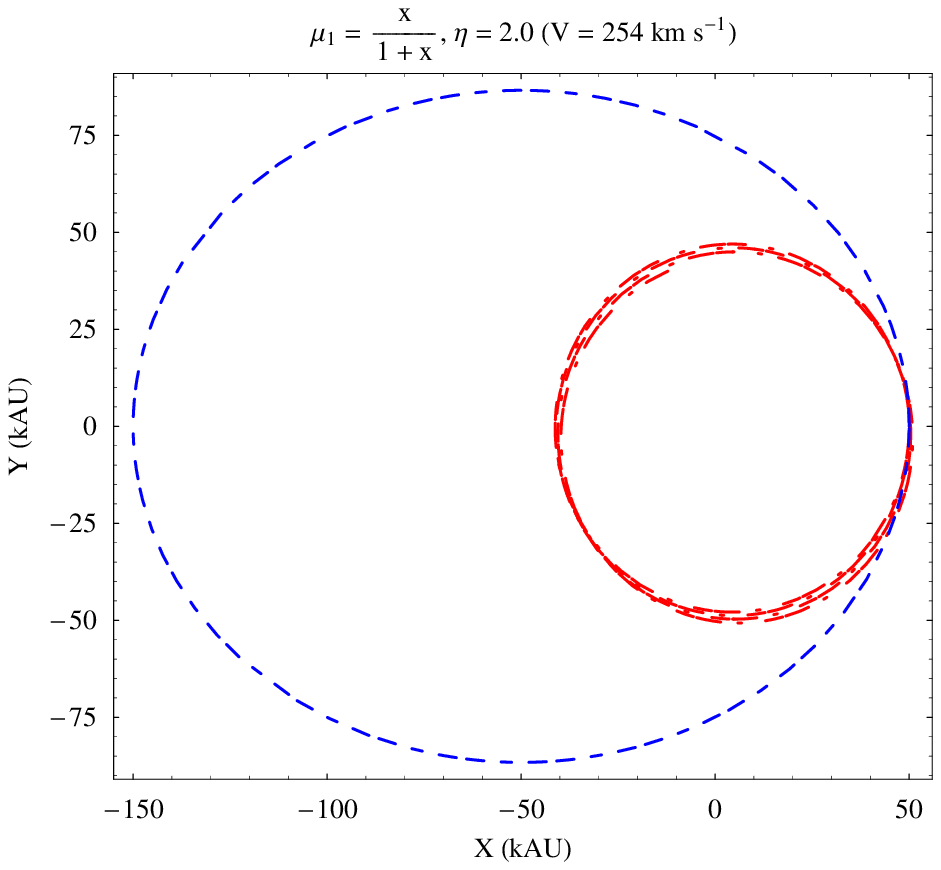,width=\columnwidth}}
\vspace*{8pt}
\caption{Numerically integrated orbits of an Oort comet with $a=100$ kAU, $e=0.5$, $P_{\rm b}=31.6$ Myr. Dashed blue line: Newton. Dash-dotted red line: MOND with $\mu_{1}$, $\eta=2.0$ ($V=254$ km s$^{-1}$). The initial conditions are $x_0=a(1-e), y_0=z_0=0,\dot x_0=0,\dot y_0=n a \sqrt{\rp{1+e}{1-e}},\dot z_0=0$. The time span of the integration is $-P_{\rm b}\leq t\leq 0$.}\label{perigal1}
\end{figure}

\begin{figure}[ht!]
 \centerline{\psfig{file=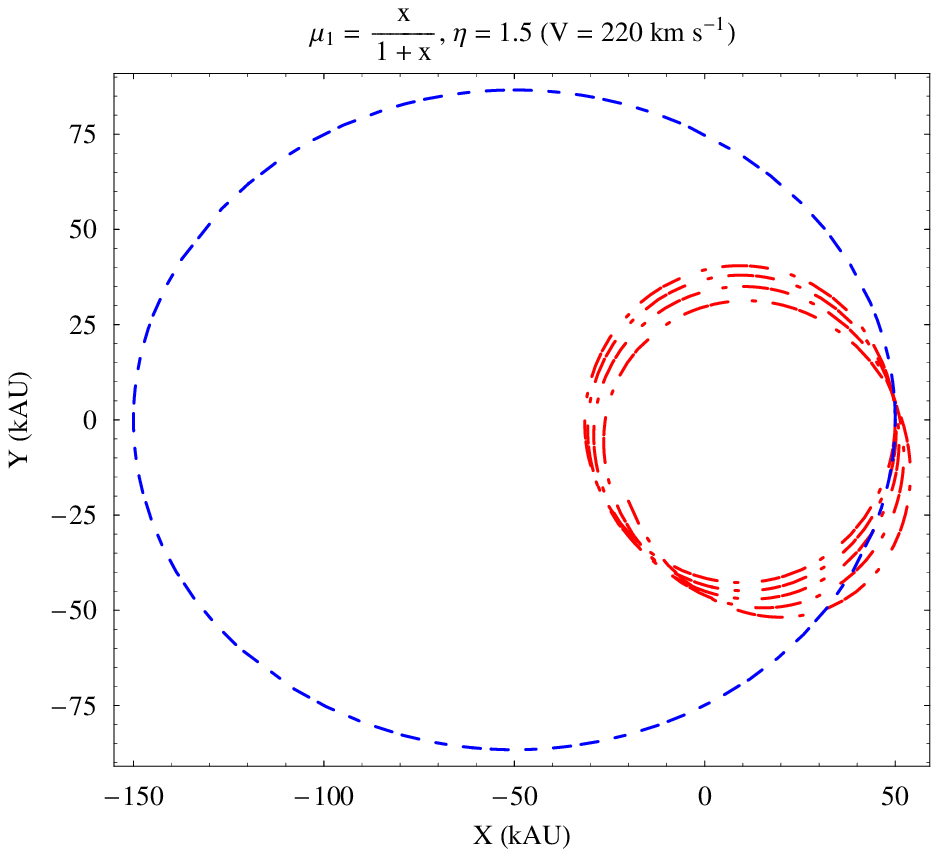,width=\columnwidth}}
\vspace*{8pt}
\caption{Numerically integrated orbits of an Oort comet with $a=100$ kAU, $e=0.5$, $P_{\rm b}=31.6$ Myr. Continuous blue line: Newton. Dash-dotted red line: MOND with $\mu_{1}$, $\eta=1.5$ ($V=220$ km s$^{-1}$). The initial conditions are $x_0=a(1-e), y_0=z_0=0,\dot x_0=0,\dot y_0=n a \sqrt{\rp{1+e}{1-e}},\dot z_0=0$. The time span of the integration is $-P_{\rm b}\leq t\leq 0$.}\label{perigal2}
\end{figure}

\begin{figure}[ht!]
 \centerline{\psfig{file=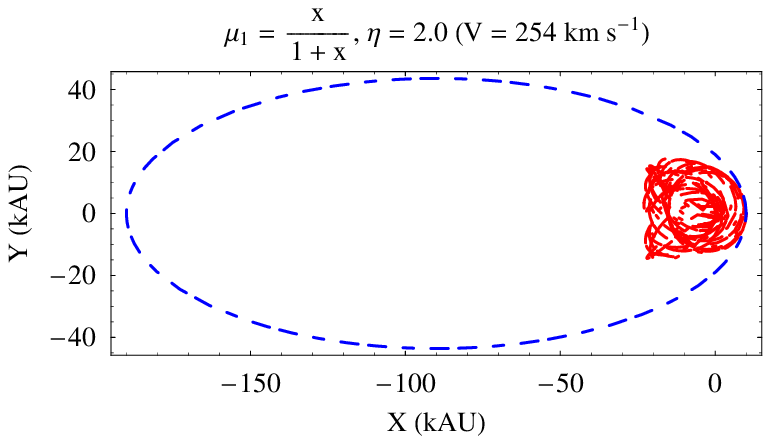,width=\columnwidth}}
\vspace*{8pt}
\caption{Numerically integrated orbits of an Oort comet with $a=100$ kAU, $e=0.9$, $P_{\rm b}=31.6$ Myr. Dashed blue line: Newton. Dash-dotted red line: MOND with $\mu_{1}$, $\eta=2.0$ ($V=254$ km s$^{-1}$). The initial conditions are $x_0=a(1-e), y_0=z_0=0,\dot x_0=0,\dot y_0=n a \sqrt{\rp{1+e}{1-e}},\dot z_0=0$. The time span of the integration is $-P_{\rm b}\leq t\leq 0$.}\label{perigal3}
\end{figure}

\begin{figure}[ht!]
 \centerline{\psfig{file=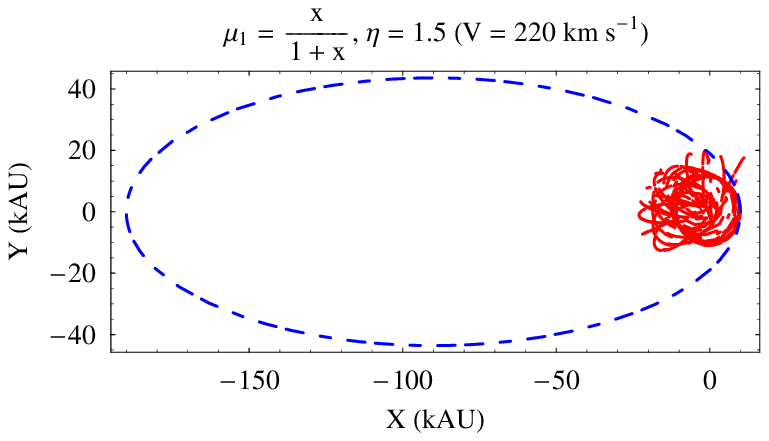,width=\columnwidth}}
\vspace*{8pt}
\caption{Numerically integrated orbits of an Oort comet with $a=100$ kAU, $e=0.9$, $P_{\rm b}=31.6$ Myr. Dashed blue line: Newton. Dash-dotted red line: MOND with $\mu_{1}$, $\eta=1.5$ ($V=220$ km s$^{-1}$). The initial conditions are $x_0=a(1-e), y_0=z_0=0,\dot x_0=0,\dot y_0=n a \sqrt{\rp{1+e}{1-e}},\dot z_0=0$. The time span of the integration is $-P_{\rm b}\leq t\leq 0$.}\label{perigal4}
\end{figure}

Let us, now, examine the case $\mu_1$.
In Figure \ref{perigal1} we show the  trajectory due to it of the Oort comet with $e=0.5$ over  $P_{\rm b}$ for $\eta=2.0$ which implies
\begin{eqnarray}
  \mu_g &=& 0.67 \\
  L_g &=& 0.33.
\end{eqnarray}
The case $\eta=1.5$, yielding
\begin{eqnarray}
  \mu_g &=& 0.60 \\
  L_g &=& 0.39,
\end{eqnarray}
is shown in Figure \ref{perigal2} for $-P_{\rm b}\leq t\leq 0$.
The case of highly elliptic orbits ($e=0.9$) is more intricate, as shown by Figure \ref{perigal3} and Figure \ref{perigal4}.
Indeed, the MOND paths resemble confuse clews confined within small spatial regions shifted with respect to the $\mu_{3/2}$ case.

\begin{figure}[ht!]
 \centerline{\psfig{file=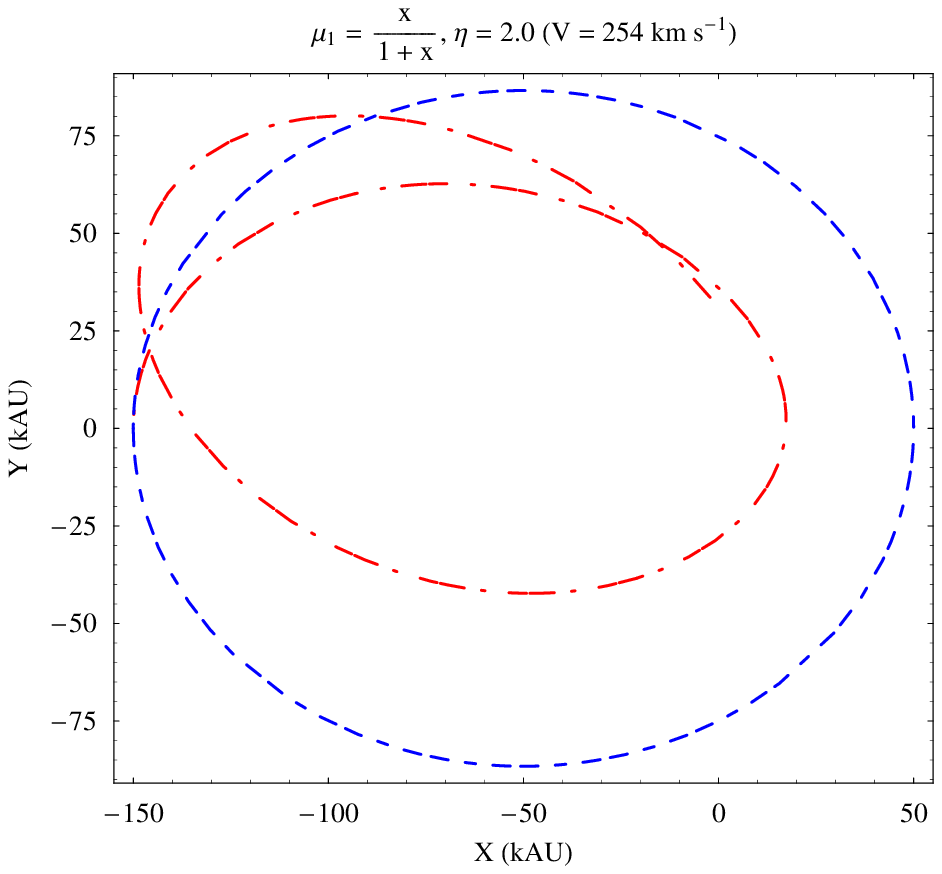,width=\columnwidth}}
\vspace*{8pt}
\caption{Numerically integrated orbits of an Oort comet with $a=100$ kAU, $e=0.5$, $P_{\rm b}=31.6$ Myr. Dashed blue line: Newton. Dash-dotted red line: MOND with $\mu_{1}$, $\eta=2.0$ ($V=254$ km s$^{-1}$). The initial conditions are $x_0=-a(1+e), y_0=z_0=0,\dot x_0=0,\dot y_0=-n a \sqrt{\rp{1-e}{1+e}},\dot z_0=0$. The time span of the integration is $-P_{\rm b}\leq t\leq 0$.}\label{apogal1}
\end{figure}

\begin{figure}[ht!]
 \centerline{\psfig{file=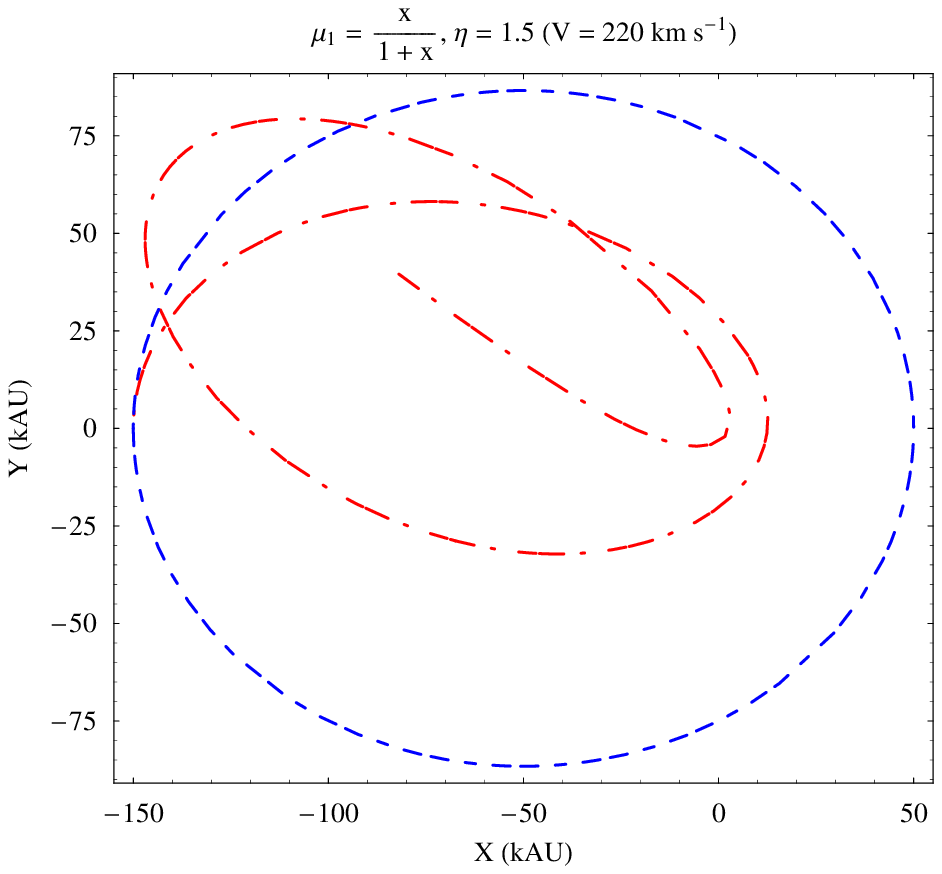,width=\columnwidth}}
\vspace*{8pt}
\caption{Numerically integrated orbits of an Oort comet with $a=100$ kAU, $e=0.5$, $P_{\rm b}=31.6$ Myr. Continuous blue line: Newton. Dash-dotted red line: MOND with $\mu_{1}$, $\eta=1.5$ ($V=220$ km s$^{-1}$). The initial conditions are $x_0=-a(1+e), y_0=z_0=0,\dot x_0=0,\dot y_0=-n a \sqrt{\rp{1-e}{1+e}},\dot z_0=0$. The time span of the integration is $-P_{\rm b}\leq t\leq 0$.}\label{apogal2}
\end{figure}

\begin{figure}[ht!]
 \centerline{\psfig{file=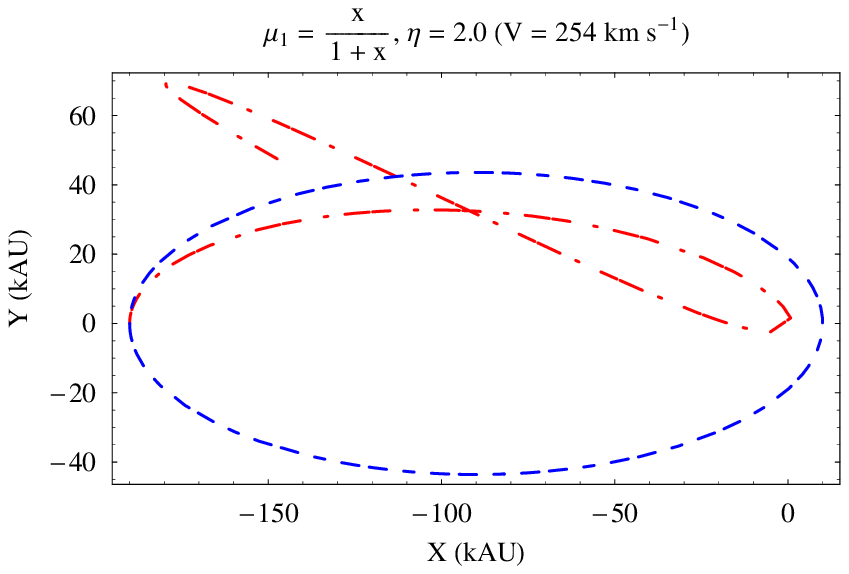,width=\columnwidth}}
\vspace*{8pt}
\caption{Numerically integrated orbits of an Oort comet with $a=100$ kAU, $e=0.9$, $P_{\rm b}=31.6$ Myr. Dashed blue line: Newton. Dash-dotted red line: MOND with $\mu_{1}$, $\eta=2.0$ ($V=254$ km s$^{-1}$). The initial conditions are $x_0=-a(1+e), y_0=z_0=0,\dot x_0=0,\dot y_0=-n a \sqrt{\rp{1-e}{1+e}},\dot z_0=0$. The time span of the integration is $-P_{\rm b}\leq t\leq 0$.}\label{apogal3}
\end{figure}

\begin{figure}[ht!]
 \centerline{\psfig{file=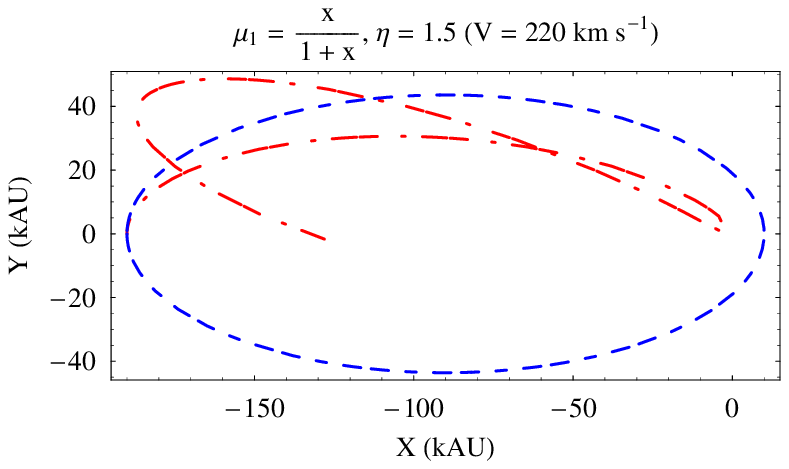,width=\columnwidth}}
\vspace*{8pt}
\caption{Numerically integrated orbits of an Oort comet with $a=100$ kAU, $e=0.9$, $P_{\rm b}=31.6$ Myr. Dashed blue line: Newton. Dash-dotted red line: MOND with $\mu_{1}$, $\eta=1.5$ ($V=220$ km s$^{-1}$). The initial conditions are $x_0=-a(1+e), y_0=z_0=0,\dot x_0=0,\dot y_0=-n a \sqrt{\rp{1-e}{1+e}},\dot z_0=0$. The time span of the integration is $-P_{\rm b}\leq t\leq 0$.}\label{apogal4}
\end{figure}

Also the trajectories of Figure \ref{perigal1}-Figure \ref{perigal4} start from the (Newtonian) perihelia. If, instead, we choose the (Newtonian) aphelia quite different paths occur also for this form of $\mu$. Figure \ref{apogal1}-Figure \ref{apogal4} show them.

\subsection{Nearly polar orbits}
\begin{figure}[ht!]
 \centerline{\psfig{file=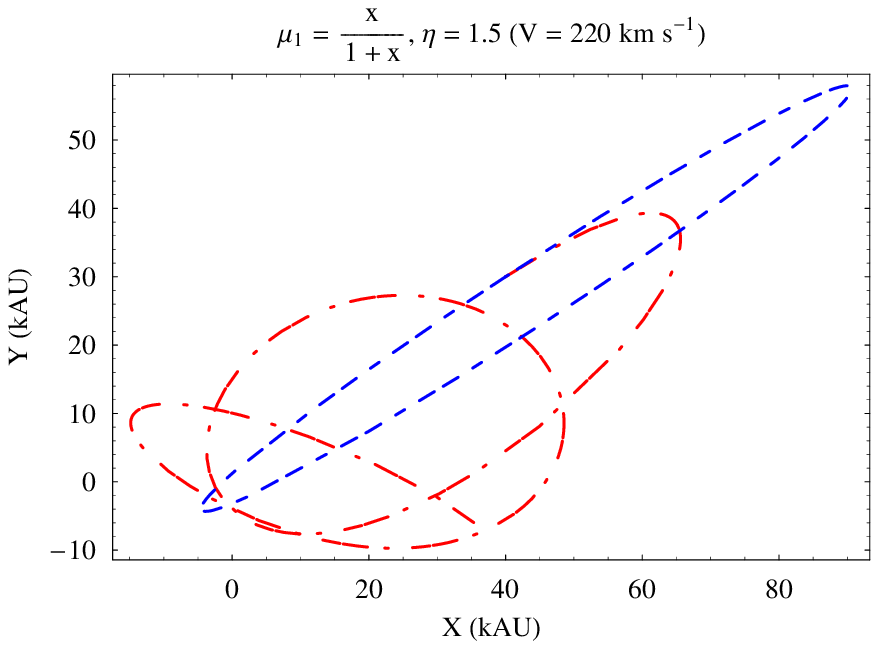,width=\columnwidth}}
\vspace*{8pt}
\caption{Section in the coordinate $\{\rm XY\}$ plane of the numerically integrated orbits of an Oort comet with $a=66.6$ kAU, $e=0.92$, $I=81$ deg. Dashed blue line: Newton. Dash-dotted red line: MOND with $\mu_{1}$, $\eta=1.5$ ($V=220$ km s$^{-1}$). The initial conditions are $x_0=40\ {\rm kAU}, y_0=30\ {\rm kAU}, z_0=5\ {\rm kAU},\dot x_0=-23\ {\rm kAU\ Myr^{-1}},\dot y_0=-15\ {\rm kAU\ Myr^{-1}},\dot z_0=-15\ {\rm kAU\ Myr^{-1}}$. The time span of the integration is $-P_{\rm b}\leq t\leq 0$.}\label{kazza1}
\end{figure}

\begin{figure}[ht!]
 \centerline{\psfig{file=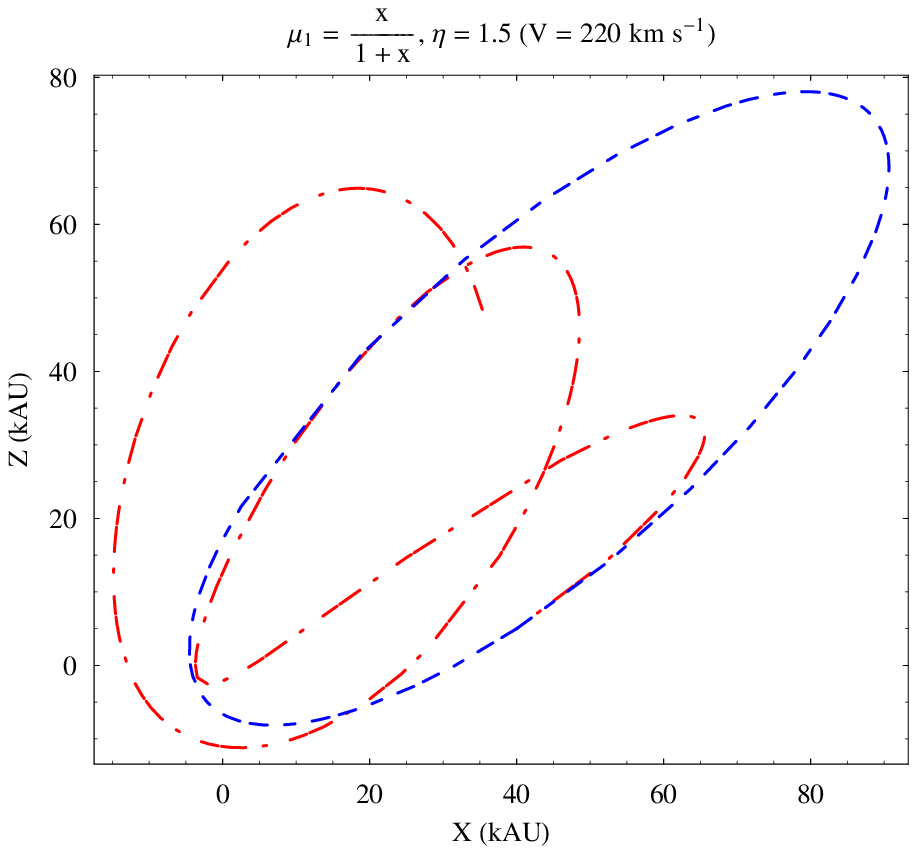,width=\columnwidth}}
\vspace*{8pt}
\caption{Section in the coordinate $\{\rm XZ\}$ plane of the numerically integrated orbits of an Oort comet with $a=66.6$ kAU, $e=0.92$, $I=81$ deg. Dashed blue line: Newton. Dash-dotted red line: MOND with $\mu_{1}$, $\eta=1.5$ ($V=220$ km s$^{-1}$). The initial conditions are $x_0=40\ {\rm kAU}, y_0=30\ {\rm kAU}, z_0=5\ {\rm kAU},\dot x_0=-23\ {\rm kAU\ Myr^{-1}},\dot y_0=-15\ {\rm kAU\ Myr^{-1}},\dot z_0=-15\ {\rm kAU\ Myr^{-1}}$. The time span of the integration is $-P_{\rm b}\leq t\leq 0$.}\label{kazza2}
\end{figure}

\begin{figure}[ht!]
 \centerline{\psfig{file=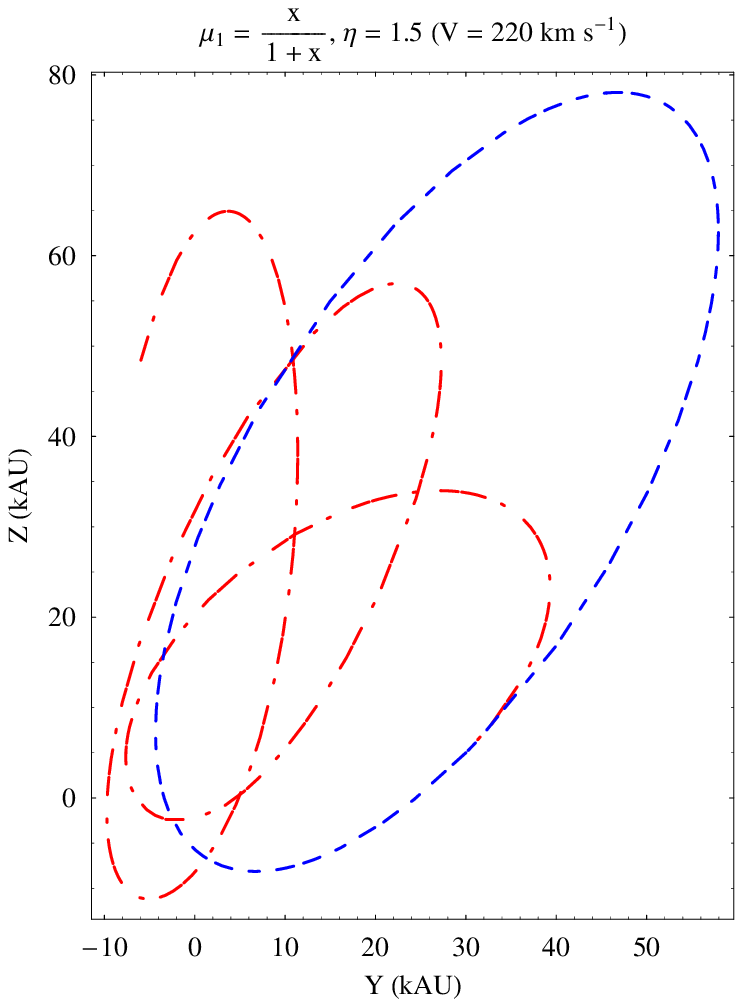,width=\columnwidth}}
\vspace*{8pt}
\caption{Section in the coordinate $\{\rm YZ\}$ plane of the numerically integrated orbits of an Oort comet with $a=66.6$ kAU, $e=0.92$, $I=81$ deg. Dashed blue line: Newton. Dash-dotted red line: MOND with $\mu_{1}$, $\eta=1.5$ ($V=220$ km s$^{-1}$). The initial conditions are $x_0=40\ {\rm kAU}, y_0=30\ {\rm kAU}, z_0=5\ {\rm kAU},\dot x_0=-23\ {\rm kAU\ Myr^{-1}},\dot y_0=-15\ {\rm kAU\ Myr^{-1}},\dot z_0=-15\ {\rm kAU\ Myr^{-1}}$. The time span of the integration is $-P_{\rm b}\leq t\leq 0$.}\label{kazza3}
\end{figure}

\begin{figure}[ht!]
 \centerline{\psfig{file=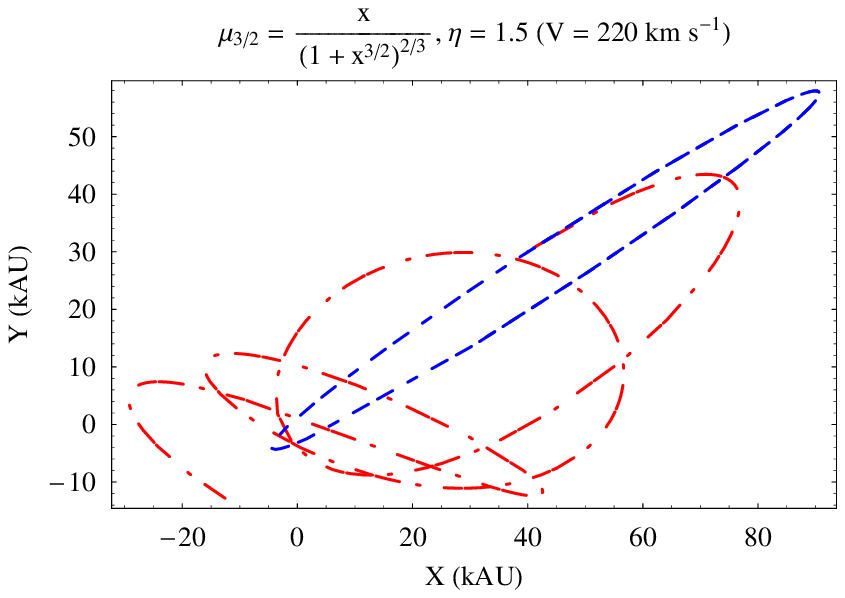,width=\columnwidth}}
\vspace*{8pt}
\caption{Section in the coordinate $\{\rm XY\}$ plane of the numerically integrated orbits of an Oort comet with $a=66.6$ kAU, $e=0.92$, $I=81$ deg. Dashed blue line: Newton. Dash-dotted red line: MOND with $\mu_{3/2}$, $\eta=1.5$ ($V=220$ km s$^{-1}$). The initial conditions are $x_0=40\ {\rm kAU}, y_0=30\ {\rm kAU}, z_0=5\ {\rm kAU},\dot x_0=-23\ {\rm kAU\ Myr^{-1}},\dot y_0=-15\ {\rm kAU\ Myr^{-1}},\dot z_0=-15\ {\rm kAU\ Myr^{-1}}$. The time span of the integration is $-P_{\rm b}\leq t\leq 0$.}\label{pippa1}
\end{figure}

\begin{figure}[ht!]
 \centerline{\psfig{file=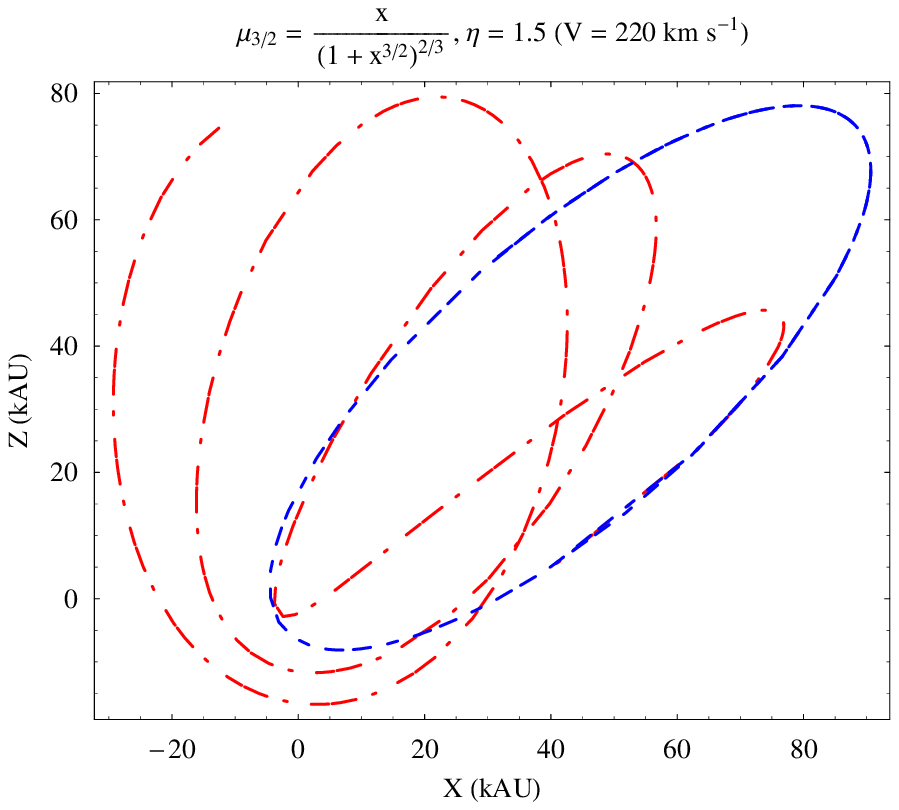,width=\columnwidth}}
\vspace*{8pt}
\caption{Section in the coordinate $\{\rm XZ\}$ plane of the numerically integrated orbits of an Oort comet with $a=66.6$ kAU, $e=0.92$, $I=81$ deg. Dashed blue line: Newton. Dash-dotted red line: MOND with $\mu_{3/2}$, $\eta=1.5$ ($V=220$ km s$^{-1}$). The initial conditions are $x_0=40\ {\rm kAU}, y_0=30\ {\rm kAU}, z_0=5\ {\rm kAU},\dot x_0=-23\ {\rm kAU\ Myr^{-1}},\dot y_0=-15\ {\rm kAU\ Myr^{-1}},\dot z_0=-15\ {\rm kAU\ Myr^{-1}}$. The time span of the integration is $-P_{\rm b}\leq t\leq 0$.}\label{pippa2}
\end{figure}

\begin{figure}[ht!]
 \centerline{\psfig{file=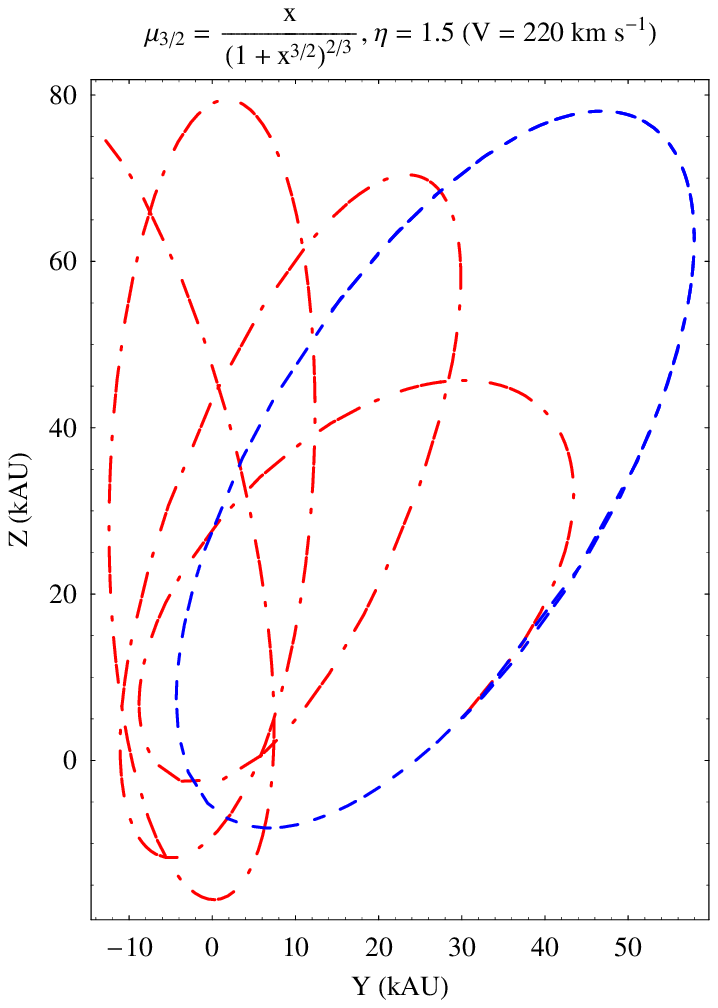,width=\columnwidth}}
\vspace*{8pt}
\caption{Section in the coordinate $\{\rm YZ\}$ plane of the numerically integrated orbits of an Oort comet with $a=66.6$ kAU, $e=0.92$, $I=81$ deg. Dashed blue line: Newton. Dash-dotted red line: MOND with $\mu_{3/2}$, $\eta=1.5$ ($V=220$ km s$^{-1}$). The initial conditions are $x_0=40\ {\rm kAU}, y_0=30\ {\rm kAU}, z_0=5\ {\rm kAU},\dot x_0=-23\ {\rm kAU\ Myr^{-1}},\dot y_0=-15\ {\rm kAU\ Myr^{-1}},\dot z_0=-15\ {\rm kAU\ Myr^{-1}}$. The time span of the integration is $-P_{\rm b}\leq t\leq 0$.}\label{pippa3}
\end{figure}

Let us, now consider the case of orbits showing high inclinations $I$ to the ecliptic.
For practical reasons, here we will only show some cases. In Figure \ref{kazza1}-Figure \ref{kazza3} we depict the sections in the coordinate planes of an orbit with $a=66.6$ kAU, $e=0.92$, $I=81$ deg for $\mu_1$ and $\eta=1.5$. The initial conditions chosen are arbitrary in the sense that, contrary to Section \ref{eclitt},  neither the perihelion nor the aphelion have been used as starting points.
The case of $\mu_{3/2}$ and $\eta=1.5$ is illustrated in Figure \ref{pippa1}- Figure \ref{pippa3}.
\section{Consequences on the Oort cloud}
The features of the MOND trajectories shown in the previous pictures suggest that EFE may have consequences on the interaction of the Oort-like objects  with passing stars \cite{Oo} by changing their perturbing effects and, thus, also altering the number of long-period comets launched into  the inner regions of the solar system and the number of comets left in the cloud throughout its history. Indeed, in the standard picture, the comets moving along very (Newtonian) elongated orbits may come relatively close to a star of mass $M_{\star}$ suffering  a  change in velocity $\Delta v$ which
approximately is \cite{Oo}
\eqi \Delta v = \rp{2GM_{\star}}{v_{\star}d},\eqf
where $v_{\star}$ is the star's velocity with respect to the Sun and $d$ is the distance of closest approach with the Oort object. Moreover, also the perturbing effects of the Galactic tides would be altered. In particular, for those particular initial configurations yielding highly shrunken paths with respect to the Newtonian case the perturbing effects of the passing stars and of the Galactic tides may get reduced. Anyway, it is very difficult to realistically predict the modifications that the Oort cloud would undergo under the action of MOND with EFE. Extensive  numerical simulations like, e.g., the one in Ref.~\cite{Masi} performed in a different context,
would be required;  they are beyond the scope of the present paper.
\section{Summary and conclusions}\lb{conclu}
We investigated the orbital motions of test particles according to MOND with EFE in  the Oort cloud ($r\approx 50-150$ kAU).
As MONDian interpolating function $\mu(x)$, we extensively used the forms $\mu_1=1/(1+x),\mu_2=x/\sqrt{1+x^2},\mu_{3/2}=x/(1+x^{3/2})^{2/3}$.
We numerically integrated both the MOND and the Newtonian equations of motion in Cartesian coordinates sharing the same initial conditions backward in time over one (Keplerian) orbital period. We considered both ecliptic and nearly polar trajectories, all corresponding to  high (Newtonian) eccentricities ($e=0.5-0.9$). In order to evaluate the characteristic MONDian EFE parameters $\mu_g$ and $L_g$ entering the problem, we used two different values ($V=220$ km s$^{-1}$ and $V=254$ km s$^{-1}$) of the circular speed of the solar system's motion through the Galaxy yielding the Milky Way's gravitational field at the Sun's location.

Our results show that the MOND orbits are quite different with respect to the Newtonian ones; in general, they are open trajectories which do not exhibit any regular patterns. Moreover, they are highly sensitive to the initial conditions in the sense that different sets of state vectors yielding the same Keplerian ellipses generate quite different MONDian paths. For certain initial configurations in the ecliptic, corresponding to Newtonian perihelion passages of highly eccentric orbits, the MOND trajectories reduce to intricate clews spanning very small spatial regions with respect to the Newtonian case. For other initial configurations, both in the ecliptic and outside it, the MOND trajectories are completely different: among other things, their spatial extensions are larger than in the ecliptic perihelion cases. As a consequence,  the structure and the dynamical history of the Oort cloud, in the deep MONDian regime, would be altered with respect to the standard Newtonian picture in a way which is difficult to realistically predict.

As further extensions of the present work, which are outside its scope, we suggest that
extensive numerical simulations of the dynamics of the Oort cloud in MOND with EFE would be helpful. Moreover, an analysis based on standard techniques of classical dynamics able to display regularities in orbits, like, e.g., surface of section,
may be fruitfully implemented.
Finally, let us note that a further issue which may be the subject of future investigations consists of the following. The numerically integrated MONDian trajectories with EFE may come close the Sun enough to fall in a Newtonian or quasi-Newtonian regime. If so, their further evolution should be obtained with a new integration of the Newtonian equations of motion starting from such new initial conditions. Consequently, the pattern of the Oort cloud may further change.

\section*{Acknowledgments}
I thank some anonymous referees for their important and constructive remarks which helped in improving this manuscript.

\end{document}